# The Dark Side of the Universe

A. De Rújula [a, b]

[a] *Instituto de Física Teórica (UAM/CSIC), Autónoma Univ. of Madrid, Spain*
[b] *Theory Division, CERN, CH 1211, Geneva 23, Switzerland*

ABSTRACT

A view on the history and current status of dark matter and dark energy, at a fairly introductory level.

## Introduction

Giordano Bruno, in 1600 CE, was burned at the stake for having defended the existence of astronomical objects other than the ones we see: *"Così io ho parlato di infiniti mondi particolari simili alla Terra"*,[1] said he during his seven-year-long trial by the Inquisition. The controversy concerning what may exist --but we do not see-- has not abated. It is no longer so potentially lethal, to the limited extent that rational and religious beliefs have parted ways.

Not exceptionally, the Greeks were pioneers on our subject matter. Epicurus (341 BCE – 270 BCE), in a letter to Herodotus, contended that there should be an infinite number of other worlds "some like this world, others unlike it"[2]. Also ahead of the times, he allowed women to join his philosophical school.

We know from very convincing data and analyses that nearly three quarters of what the universe is made of is in the form of a rather mysterious "dark energy". Moreover, we have convincing evidence that about four fifths of the matter in the universe is also "dark", and different from anything we know. "Dark" in this sense means neither emitting or absorbing electromagnetic radiations such as light. The cited amounts of the different substances refer to their average contribution to the energy density of the universe.

---

[1] Thus have I spoken about infinitely many particular (concrete?) worlds similar to ours.
[2] It is not my intention to write a history of dark substances. An excellent source, with a very complete list of references, is G. Bertone and D. Hooper, "A history of Dark Matter", arXiv:1605.04909, wherein I learned a

lot.

The situation I have described has no precedent: what we know that we do not know has never been so "abundant". The challenge to the physical sciences is mind-boggling. Excellent times to be a scientist.

## The first scientific steps

Looking at the sky and trying to interpret the nature and motions of celestial bodies were probably amongst the first things that distinguished humanoids from other species. Looking, eventually with telescopes, continued for a long time to be the only way to make progress in astronomy. That changed radically with Newton's discovery of the laws of gravitation. In particular it became possible to "see" something "dark".

A most spectacular gravity-aided discovery was that of Neptune, though it was not the first[3]. In 1846 Urbain Le Verrier and John Couch Adams, adopting a suggestion of Alexis Bouvard to explain the anomalies of Uranus' orbit, predicted the existence and whereabouts of a "dark" planet. Neptune was found, one degree away from its predicted position, by John Galle and his student Heinrich Louis D'Arrest the night of the very day they were advised to look for it. That is the nice way of putting it. Whether or not Adams had truly contributed to the prediction triggered a not uncharacteristic battle in academic circles, though Adams himself acknowledged Le Verrier's priority[4]. Having been so successful Le Verrier enthusiastically proceeded to attribute the anomaly in the advance of Mercury's perihelion to the effect of yet another dark planet: *Vulcan*. It is well known that the correct interpretation of this anomaly, provided by Einstein's theory of General Relativity, is even more interesting than yet another planet. Will the current dark matter epic end like Neptune's or like Vulcan's?

A large number of eminent scientists debated for about a century the existence of "dark" matter, which during this period meant ordinary matter not in stars sufficiently luminous to be visible. Jules Henri Poincaré (who invented the expression *matière obscure*) and Lord Kelvin treated the not very distant stars in our galaxy as a gas, to relate their velocity dispersion to the total amount of gravitating matter. They concluded that the amounts of ordinary and dark matter were comparable and perhaps to be accounted for by under-luminous stars, and/or "nebulous" and "meteoric" matter.

The *Virial theorem* states that, in a stable system of objects mutually bound by gravity, the total kinetic energy of its components equals minus one half their potential energy. For galaxies in a *cluster*, for instance, one may measure their velocities, assume the values of their masses and test the theorem. If the kinetic energies exceed what the theorem would "demand", the amount of matter contributing to the gravitational potential has been underestimated: a fraction of the total matter must be dark. Poincaré may have been the first to apply this theorem to astrophysics, in his *Leçons sur les hypothèses cosmogoniques professées à la Sorbonne*, a not uncharacteristically French title.

---

[3] In 1844 Friederich Bessel argued that the peculiar proper motions of Sirius and Procyon were due to their unobserved companion stars.

[4] See https://en.wikipedia.org/wiki/Discovery_of_Neptune and references therein.

Around the 1930's a decisive character entered the long dark saga: Fritz Zwicky, a Caltech professor who had graduated in what is now the ETH[5] in Zurich. He had a harsh character and would refer to some people as "spherical bastards", for the invariance of their bastardy under rotations. He analysed data on the Coma Cluster, obtained by Edwin Hubble and Milton Humason, concluding that some galaxies had exceedingly large velocities and that "*If this would be confirmed, we would get the surprising result that dark matter is present in much greater amount than luminous matter*". Doubters such as Eric Holmberg argued instead that the "fast" galaxies in a cluster were not bound to it, they were temporarily visiting outsiders.

One systematic reason why the early numbers on the amount of dark matter in clusters were quite wrong is that the value of the Hubble constant (which relates cosmic distances to easily measurable redshifts) was then an order of magnitude larger than its current estimates. This constant is so inconstant and its value is so passionately and interminably debated that it deserves a separate chapter, a dark energy interlude in the history of dark matter.

## The Hubble plot and beyond

The very first solid evidence for the dark energy of the universe arose from an extension to much larger cosmic distances of the work of Hubble and others on the expansion of the universe. The history of the subject illustrates how difficult the interpretation of astrophysical observations is, and how it is not that easy to have astrophysicists agree, a point to be always kept in mind. Laboratory physics and even cosmology are more peaceful domains... in general.

We can estimate by eye how far a not very distant object is. Our eyes see it at slightly different angles, and our brain exploits the angular difference and the separation between our eyes to provide the estimate. To measure distances to stars a *stellar parallax* method is used, with the eye-to-eye distance substituted by the size of the Earth's orbit (and up to six months of patience). The first stellar parallax measurement was made by Friedrich Bessel in 1838 for the star 61 Cygni, 11.4 light years away. The star-mapping satellite Gaia and very long interferometry antenna arrays reach the largest distances currently measurable this way, of the order of ten kiloparsecs ($\sim 3 \times 10^4$ light years). Parallax measurements are the first and only foolproof steps in a *cosmic ladder* of successive partially overlapping measures of distance, each providing a calibration for the next, further reaching one.

*Classical Cepheids* are stars whose luminosity varies in a periodic manner. They are named after δ Cephei, discovered to be variable in 1784 by the amateur astronomer John Goodricke, FRS. Their trivially measurable periods (~ 1 to 50 days) are related to their absolute magnitude or luminosity (the amount of light a nearby "calibrating" observer at a known distance would measure). From the observed period of more distant Cepheids one can infer their absolute magnitude, which compared with the actually measured one provides an estimate of the distance. This way Cepheids become *standard candles.* "Standardizable", one ought to say, for many corrections 4must be made to homogenize the results. Here and in what follows the devil is in these details, or in the "small print" of the observer's articles.

---

[5] ETH: Eidgenössische Technische Hochschule, or Federal Institute of Technology.

A *redshift, z,* is a measure of the amount by which the wavelenghts of emitted and observed radiation --typically a given atomic line-- differ[6]. The origin of this difference may be an actual relative velocity between the emitter and the observer, the expansion of the Universe[7] (an apparent "velocity"), or a combination thereof.

In 1929 Edwin Hubble combined his measurements of distance, *d*, mainly based on Cepheids in close-by galaxies, with the redshifts (and consequent velocities, $v$) measured by Vesto Slipher and Milton Humason, obtaining a linear relation, $v = H_0 d$, with $H_0$ the Hubble "constant". This relation, unfairly called Hubble's law, was obtained by Georges Lemâitre two years earlier. $H_0$ is the measure of the rate at which the universe is currently expanding, its subindex "0" means now, the expansion rate is supposed to change with time (but be independent of direction).

Hubble obtained a value $H_0 \approx 500$ kilometers per second per Megaparsec, in the abominable units used by cosmologists for what is simply an inverse time (roughly equal to the age of the universe). Subsequent estimates have converged, to some extent, to values one order of magnitude smaller. During the second half of the XX century, two schools fiercely fought over the value of $H_0$. Gerard de Vaucouleurs and his supporters defended a value close to 100, Allan Sandage, a former student of Hubble who spent all of his life's nights at the telescope[8], was adamant: circa 50 was the answer. Once again the difference could be traced to the different choices of "standard" Cepheids[9]. In 1996 a temporary truce was reached at midway: 70 was *then* the answer[10].

Subsequent technological progress resulted in several new ways to measure $H_0$ and, as you may have guessed, we are back to the traditional controversy. There are three main new methods, involving the analysis of the anisotropies of the cosmic microwave background radiation (MWBR or CBR) and the addition of two new stellar standard candles. The first, *Type Ia supernovae* (SNIa), are binary stars, one a white dwarf, the other a companion of various types. The dwarf accretes matter from the companion till it reaches an instability and suffers a thermonuclear explosion. The details are not well understood but the result is an intense *light curve,* which can be standardized to obtain a distance estimate. The second new candle is *the tip of the red giant branch* (TRGB). Stars similar to the Sun evolve in the color-luminosity[11] diagram, tracing a path that culminates in the mentioned tip. Such stars also serve to determine the distance to their host galaxy.

---

[6] For a relative velocity, $v$, much smaller than the speed of light, $c$, $z = \lambda(ob)/\lambda(em) - 1 \approx v/c$, with $\lambda[em]$ and $\lambda[ob]$ the emitted and observed wavelengths. A negative *z* would be a blueshift.
[7] Consider ants at rest (or almost so) on the surface of a balloon that is steadily inflating. Every ant thinks the others are moving away at a velocity proportional to distance (the Hubble law!). The space *itself* of the universe is similarly expanding.
[8] Or so he told me one night, thousands of miles away from the nearest decent telescope.
[9] They differed in "metallicity": their content in elements heavier than helium.
[10] See, for instance, Sylvestre Huet, *La constante de Hubble ne varie plus*, Liberation, October 15[th] 1996.
[11] The Hertzsprung–Russell diagram is a scatter plot of stars showing the relationship between the stars' absolute magnitudes or luminosities versus their types or effective temperatures.

The recent results of measurements that include the CBR information[12], the use of Cepheids and SNIa[13], and the TRGB[14] are, respectively and in the usual units:

$$H_0 = 67.44 \pm 0.58$$
$$H_0 = 74.03 \pm 1.42$$
$$H_0 = 69.8 \pm 0.8\,(stat) \pm 1.7\,(syst)$$

The third and most recent result, supposed to resolve the earlier "tension" between the first two, fell right in the middle! Moral: we do not yet know the value of $H_0$, nor whether it depends on the redshift(s) at which it is extracted from the data (the CBR method involves much larger redshift than the others). Nihil novi *super* sole[15].

The large lower part of Figure 1 shows the satisfactory extent to which the different distance measurements of the steps of the cosmic ladder overlap and agree[13]. The upper left insert, a modern version of the Lemaître-Hubble diagram, shows the comparison of the data as a function of redshift with various cosmological models to be discussed in detail anon. This comparison offers the first evidence for an accelerated expansion of the universe[16], an evidence for which Saul Perlmutter, Brian Schmidt and Adam Riess (in that order) got the 2011 Nobel Prize *"for the discovery of the accelerating expansion of the Universe through observations of distant supernovae"*. The acceleration, as we shall discuss in detail, is attributed to the effect of the dark energy of the universe.

---

[12] https://www.cosmos.esa.int/web/planck. To extract $H_0$, the CMB data are combined with baryon acoustic oscillation results (to be later discussed), and with SN Ia data, to "descend" the ladder till our location.
[13] A. G. Riess et al., ApJ, 876, 85 (2019), A. G. Riess, Space Telescope Science Institute, Volume 37, Issue 2 (2020);
[14] Wendy L. Freedman et al., ApJ **882** 34 (2019).
[15] *Ecclesiastes* 1:9, Vulgate Bible.
[16] S. Perlmutter et al., The Astrophysical Journal, Volume **517**, 565 (1999). A. Riess et al., The Astronomical Journal **116**, 1009 (1998).

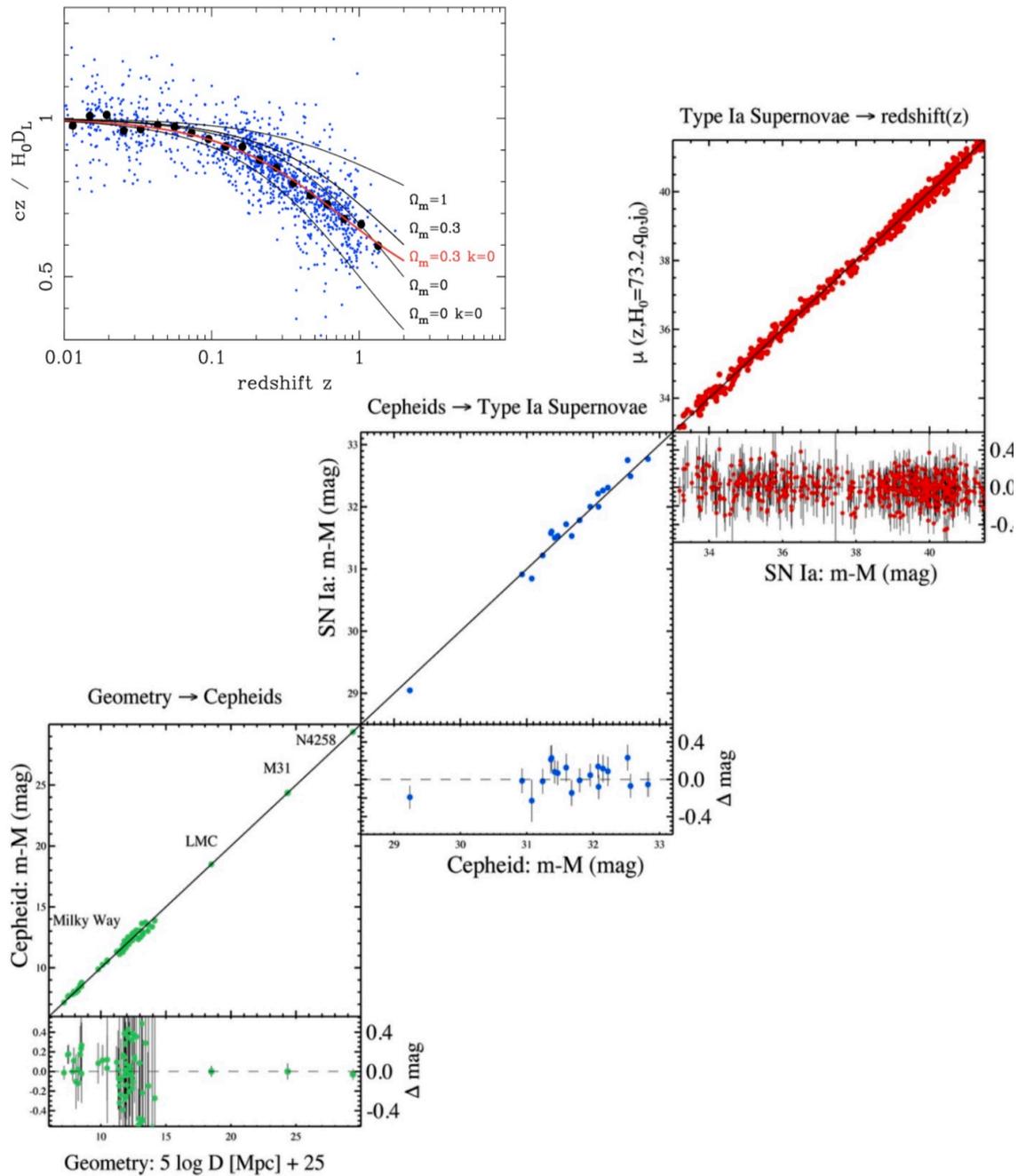

Figure 1. Upper left insert: Analysis of the SNIa data in various cosmological models. The rest: comparison between the distance measurements employed in successive steps of the parallax/Cepheid/SNIa cosmic ladder[13].

The measured $H_0$ values that I have quoted are not the only ones being recently gathered and reported. A compendium[17] is given in Fig. 2, from which one can conclude that the XXth century controversy between "schools" defending two different results is not over.

---

[17] L. Perivolaropoulos and F. Skara, arxiv:2105.05208

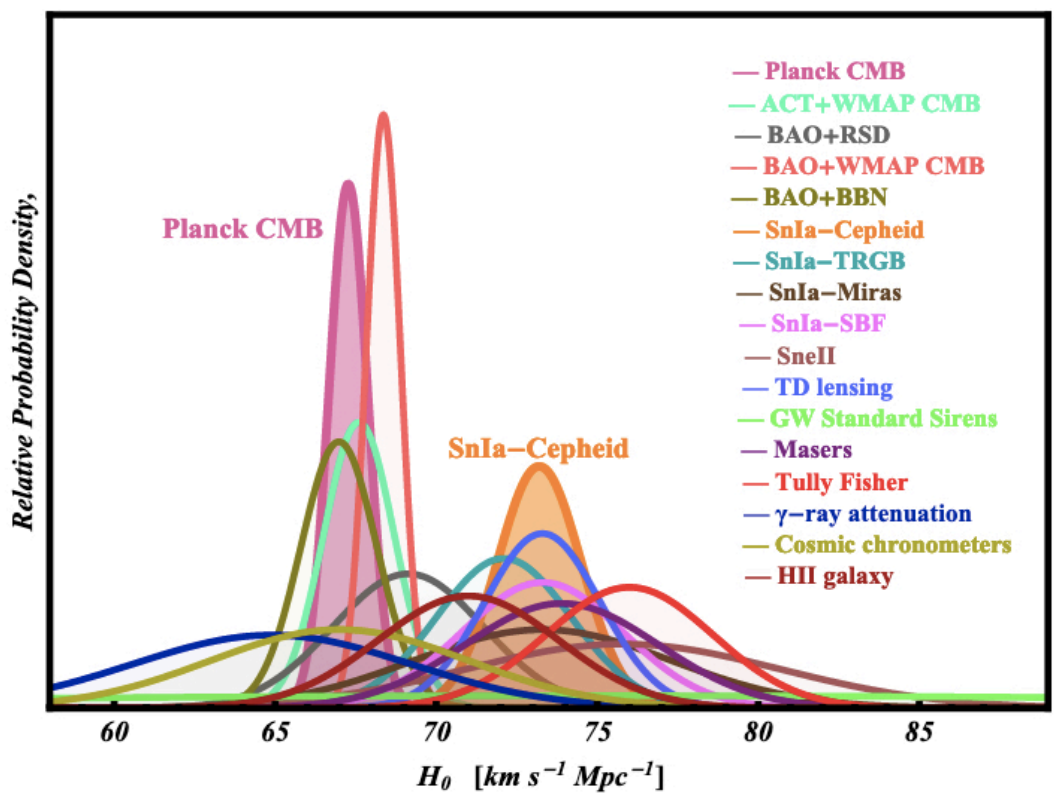

Figure 2. Ranges of values of $H_0$ extracted from various observations or combinations thereof[17].

## Dark matter in galaxies and clusters

In the 1960's astrophysicists began to seriously ponder what dark matter may be and to find what it is not[2]. Free Hydrogen in the Pegasus I cluster was found by Arno Penzias to be less than 10% of its virial mass. To study a possible ionized gas, Meekings and collaborators looked for X-ray signatures of intracluster gas in Coma, limiting it to "less than 2% of that required for gravitational binding".

A telltale clue, that of dark matter in individual galaxies, slowly ripened from 1914 to circa 1980. The main conclusion, reached with great toil and prudence, was that this dark matter was *not* ordinary matter hiding in some somber disguise, such as hydrogenic "snow" balls. The main information resided in two observations: the luminosity of a galaxy as a function of distance, *R*, from the galaxy's central region; and its *rotation curve*: the velocity *v(R)* of small amounts of ordinary, "visible" matter, in gravitationally bound orbits. As technologies improved, greater distances could be explored, with increasingly intriguing results.

In 1914 Andromeda (M31) was not yet proved by Hubble to be a (spiral) galaxy, external to ours. Yet, Slipher and Max Wolf, observing its spectral lines, concluded that it rotated around its axis. For a long time it was concluded that a constant "mass to light ratio" (no dearth of star light at large *R*) was good enough to describe the rotation curves. The development of radio astronomy following WW II added a precise new tool of observation. With it, Edward Purcell and his student Harold Ewen discovered in 1951 the 21 cm line emitted by clouds of atomic hydrogen[18] in the Galaxy, whose existence had been predicted by Christoffel van de Hulst.

During the 1970s, the observations finally made the conclusion unavoidable: galaxies contain large amounts of non-ordinary, truly novel "dark" matter. The reason was simple: in a galaxy whose mass is accounted for by matter visible up to a certain distance from its center, the rotation curve should, beyond that distance, have a velocity *v(R)* proportional to *$1/R^{1/2}$*, exactly like the planets do in their orbs around the sun, and for the same reason (Newton's laws). But the rotation curves of large numbers of galaxies are flattish: *v(R)* remains roughly constant up to distances where the signals become too weak to measure. The example of the galaxy M33 is shown in Figure 3.

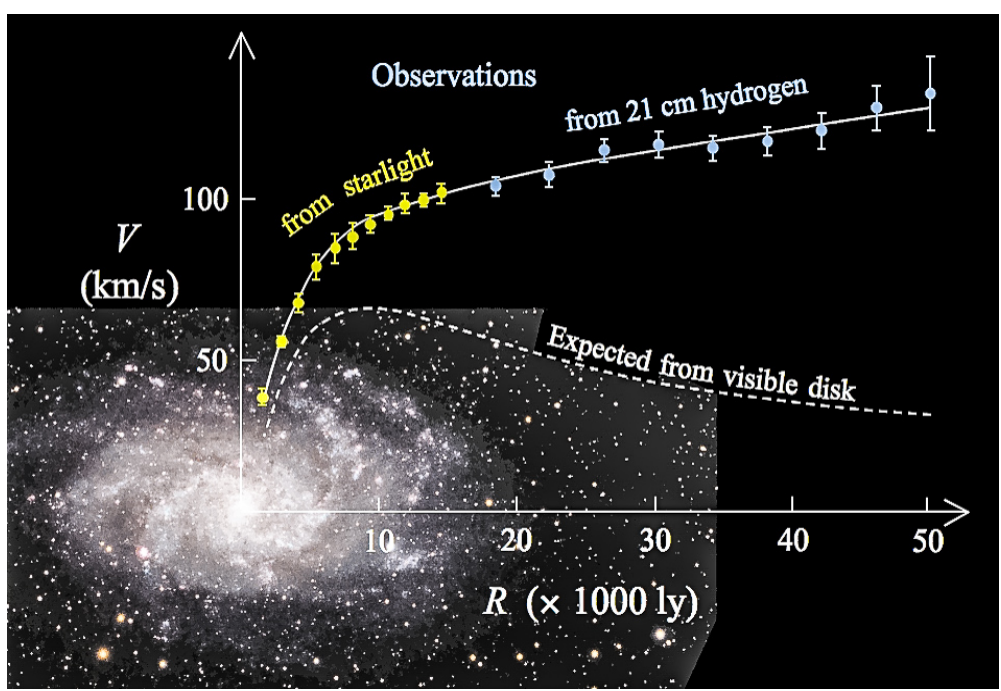

Figure 3. M33 rotation curve. The data points and the continuous line are the observed velocities. Illustration in E. Corbelli and P. Salucci, MNRAS, 311, 441 (2000).

In establishing the "flatness" of rotation curves, the work of Vera Rubin, Kent Ford[19] and collaborators was crucial and much doubted for starters[20]. Yet, as early as 1970, Ken Freeman daringly concluded: *there must be in these galaxies additional matter which is undetected, either optically or at 21 cm … its distribution must be quite different from the exponential distribution which holds for the optical galaxy*[21]. I shall not outline the very many crucial observations[2] that led to the conclusion that galaxies, small groups of galaxies and galaxy clusters contain about one order of magnitude more dark than ordinary matter. As Sandra Faber and John Gallagher put it in 1979: *after reviewing all the evidence, it is our opinion*

---

[18] 21 cm is the wavelength of the photons emitted in the transition between the two lowest-energy states of the hydrogen atom: with parallel or antiparallel spins of the electron and proton.
[19] Vera Rubin and W. Kent Ford, Jr., *Rotation of the Andromeda Nebula from a Spectroscopic Survey of Emission Regions*. The Astrophysical Journal. **159**: 379ff (1970).
[20] For a review, see V. Rubin, *A Century of Galaxy Spectroscopy*. Astrophys. J. **451**: 419ff (1995).
[21] K. C. Freeman, Astrophys. J. **160**, 811 (1970).

*that the case for invisible mass in the universe is very strong and getting stronger*[22]. This is now the consensus.

Einstein reached instant fame in May 29, 1919, when it was observed that the sun's gravitational field, acting as a convergent lens, deflected the light coming from stars that, during a solar eclipse, were visible close to its direction[23]. This was considered a triumph of Einstein's though his first prediction for the angles of deflection was wrong by a factor of two. His theory of General Relativity provides the right answer. Astrophysical objects darker than the sun also observably deflect light. To do it all they need to have is a sufficiently large mass.

**Brown dwarfs** --which are not brown-coloured-- are defined as failed stars, in the sense of being insufficiently massive to initiate the hydrogen-fusing nuclear reaction that makes the sun shine. Their masses, by (awkward) definition, lie in the range from 13 to 80 Jupiter masses, that is, 1.2% to 7.6% of a solar mass, $M_\odot$. Many brown dwarfs have been found, a bound pair of them (the Luhman 16 system) being the third starry object closest to the sun, after the Alpha Centaury triple system and Barnard's star. These under-luminous objects are dark matter candidates made of ordinary or *baryonic* matter (neutrons and protons are the "baryons" constituting ordinary matter, along with electrons).

When a sufficiently massive invisible object transits close to the line of sight to a star, the star's light is "micro-lensed", in a characteristic way, by the gravitational field of the otherwise unseen object[24]. Micro-lensing candidates include black holes, neutron stars and brown dwarfs. Not too elegantly, they are often collectively referred to by the initials of Massive Astrophysical Compact Halo Objects[25]. More elegantly, one of the main experiments that searched for them was called *Expérience pour la Recherche d'Objets Sombres*[26]. A third one is the *Optical Gravitational Lensing Experiment*[27]. Ogle, according to the Cambridge English Dictionary, means "to look at someone with obvious sexual interest".

EROS concluded that MACHOs in the mass range $0.6 \times 10^{-7} M_\odot$ to $\sim 15\, M_\odot$ are ruled out as the main constituents of the dark mass of the Galaxy. It also limited to 8% an earlier finding by the MACHO team that 40% of the dark mass of the Large Magellanic Cloud could be in MACHOS of mass in the range $\sim 0.15\, M_\odot$ to $\sim 0.9\, M_\odot$. It is not easy to summarise astrophysical findings; they may be more or less uncontroversial, but they are always controverted.

**Black holes** (BHs) require no introduction and have never been so fashionable. The ones observed until very recently had masses in two different ranges. The *stellar* ones result from the collapse of the central region of an aged massive star, which results –via a mechanism not yet understood in satisfactory detail-- in the ejection of the outer layers of the star, that shine as a "core-collapse" supernova[28]. The resulting stellar BHs have masses from a few to a few tens of $M_\odot$. The supermassive ones observed at the centres of many galaxies range in the millions or billions of solar masses. Only the stellar black holes have a well-established origin.

---

[22] S.M. Faber and J. S. Galagher, ARA&A **17**, 135 (1979).
[23] *Lights All Askew in the Heavens*, The New York Times, November 10, 1919.
[24] B. Paczynski, Astrophys. J.304, 1 (1986);
[25] MACHO experiment: C. Alcock et al., ApJ, 542, 281 (2000); superMACHO:
*The superMACHO microlensing survey, Impact of Gravitational Lensing on Cosmology*; Proceedings IAU Symposium No. 225, 2004, Y. Mellier, & G. Meylan, eds, https://arxiv.org/abs/astro-ph/0409167
[26] P. Tisserand et al. [EROS-2 Collaboration], Astron. Astrophys. 469, 387 (2007) [astro-ph/0607207], http://eros.in2p3.fr/; Anne M. Green, Phys. Rev. **D96**, 043020 (2017), arxiv:1705:10818
[27] P. Mroz et al. ApJS, 244, 29 (2019); arXiv:1906.02210v2

*Intermediate mass* BHs are defined to have masses in the range $\sim 10^2$ to $10^5 M_\odot$. The first ever to be clearly observed, by the LIGO/Virgo gravitational-wave detectors, was the result of a BH/BH merger dubbed GW190521, for the detection's date[29]. Its measured mass was "intermediate": $142^{+28}_{-16} M_\odot$. The fusing BHs had masses $85^{+21}_{-14} M_\odot$ and $66^{+17}_{-18} M_\odot$. This is most intriguing, for theorists contend[30] (or used to contend) that there is no mechanism for the stellar birth of a BH with $50\, M_\odot \leq M \leq 140\, M_\odot$[31]. At least one of the fusing BHs was not supposed to exist. A boring explanation is that these "forbidden" objects are the result of previous BH/BH fusions. More excitingly, they may be *primordial,* that is, born well before the stars[32].

*Light* BHs of mass less than $\sim 1\, M_\odot$ are not supposed to be made by stellar processes. If discovered, they would be excellent candidates to be primordial[33].

We shall eventually get to discuss the current standard model of the universe. But much before it was developed, Boris Yakovlevich Zel'dovich and Igor Dmitriyevich Novikov, used the "hot cosmological model" of the very early universe to argue that sufficiently large local density fluctuations in the primordial "soup" would result in black holes[34]. In the current "inflationary" view of the universe there were no baryons (yet) in the universe when primordial BHs would have been born. Thus they are potential candidates for non-baryonic dark matter. Stellar BHs lose their baryon number information to an external observer but they are "baryonic" by birth.

Can black holes be a substantial component of dark matter? The answers range from the statement that this is only possible for BHs in two narrow mass ranges: $10^{-16}$ to $10^{-11} M_\odot$ and $10^{14}$ to $10^{19} M_\odot$[35] to the brave contention that all of dark matter could be in primordial black holes in mass ranges including the one observed via gravitational waves[36]. This is currently a subject of intense discussion and can only be summarized by saying that nobody knows "for sure" how primordial black holes may have been born, their mass and space distributions or their "biography".

---

[28] The same mechanism, in less massive stars, gives rise to neutron star remnants.

[29] For relatively recent results, see R. Abbot et al., arXiv:2010.14533

[30] K. Belczynski et al., A&A, **594**, A97 (2016) is an early reference to the lower limit.

[31] This is called the *pair-instability* gap. The formation of electron-positron pairs is supposed to prevent collapse into a BH of mass in the cited range and, for higher stellar and resulting BH masses, to trigger a pair-instability supernova.

[32] A possibility revived by the detection of gravitational waves, see e.g. S. Bird, et al., Phys. Rev. Lett. **116**, 201301 (2016), arXiv:1603.00464. S. Clesse and J. García-Bellido, Phys. Dark Univ. **15**, 142 (2017), arXiv:1603.05234. M. Sasaki et al., Phys. Rev. Lett. **117**, 061101 (2016) [erratum: arXiv:1603.08338].

[33] G. F. Chapline, Nature, 253, 251 (1975).

[34] Y. B. Zel'dovitch and I. D. Novikov, *The Hypothesis of Cores Retarded During Expansion and the Hot Cosmological Model*. Soviet Astronomy, **10,** 602 (1966). In the West it is more common to quote B. J. Carr and S. W. Hawking, MNRAS **168**, 399 (1974).

[35] This strong statement is weakened for BH mass distributions not assumed to be narrowly peaked. See, e.g. B. Carr and F. Kühnel, arXiv:2006.02838v3 and tens of references therein.

[36] B. Carr, S. Clesse, J. García-Bellido and F. Kühnel, Phys. Dark Univ. 31, 100755 (2011), arXiv:190608217v3.

In a very recent and detailed review of constraints on BHs, primordial or not, Bernard Carr and collaborators[37] conclude that they cannot provide more than ~1/10 of the observed dark matter abundance (and much less in certain mass ranges). This is except in an unconstrained range of $10^{-15}$ to $10^{-10}$ $M_\odot$, and comes with the proviso that --as is generally the case in astrophysics-- the conclusions depend to various degrees on theoretical assumptions.

An extreme example of the *gravitational lensing* phenomenon takes place when the observer, the lens and the light source are very precisely aligned. In that case the observed image is a ring, usually called an *Einstein ring*, (not) to honor Orest Chwolson, who invented the concept twelve years before Einstein. The deflection angle is proportional to the lensing mass. This way the masses of galaxies and galaxy clusters have been observed to exceed by an order of magnitude their ordinary-matter mass. Some such rings are shown in Figure 4.

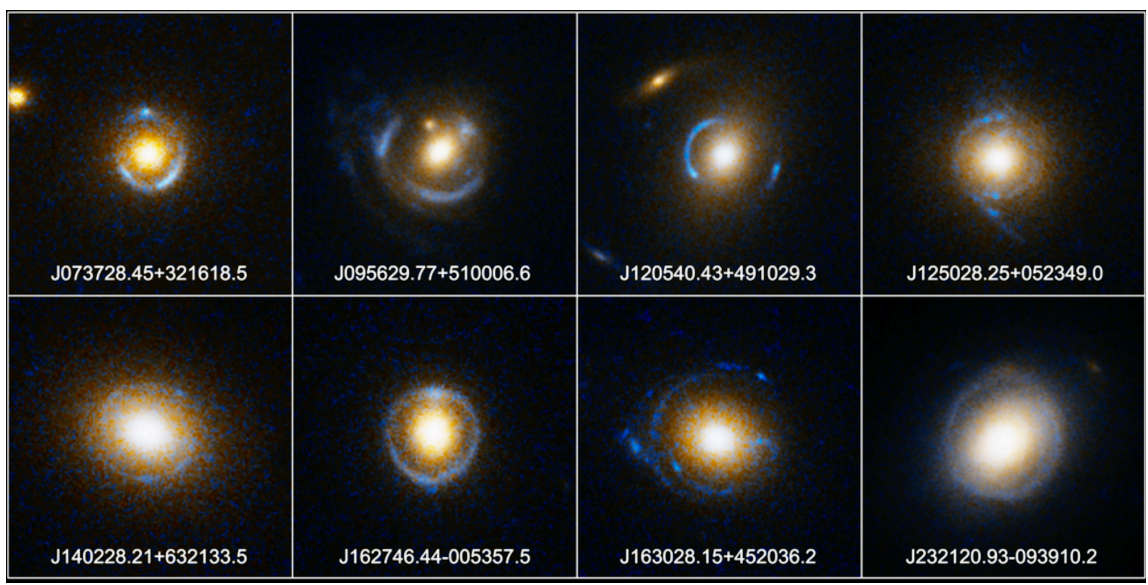

Figure 4. Chwolson rings observed with the Hubble Space Telescope[38].

The estimated amount of dark matter in galaxies and clusters is consistent with all or most of the dark matter in the universe. To ascertain how much there actually is one must look at it (the universe).

## Dark matter aplenty

The distance versus redshift measurements we have mentioned imply that the universe is expanding and was denser and hotter in the past. One would not accept such a sweeping notion without testing its predictions. The most impressive ones are the existence of the **Cosmic Background Radiation** (CBR, or MWBR) and the relative abundances of the *primordial elements*: hydrogen and a few light

---

[37] Bernard Carr, Kazunori Kohri, Yuuiti and Jun'ichi Yokoyama, arxiv:2002:12778.
[38] NASA/ESA/SLACS Survey Team: A. Bolton (Harvard/Smithsonian), S. Burles (MIT), L. Koopmans (Kapteyn), T.Treu (UCSB), L. Moustakas (JPL/Caltech).

chemical elements that are not the result of later fusion processes in stars.

When the universe was younger than $t_R \sim 380000$ years[39] (redshift $z_R \sim 1100$) its temperature was high enough for atoms to be ionized; there were protons and electrons, but no hydrogen. Photons were in thermal equilibrium with charged particles and their mean free paths were short. As time elapsed beyond $t_R$ protons and the nuclei of the primordial elements combined with electrons and became atoms. The previous plasma became a gas, transparent to the preexisting photons, then at a temperature $T_R \sim 3000K$. By now these "liberated" photons are redshifted by a factor $z_R$ to their current temperature of $T_0 = (2.72548 \pm 0.00057)$ K, in the microwave range[40]. The CBR has a current number density of $n_\gamma \sim 400$ photons/cm$^3$.

The CBR was serendipitously discovered in 1964 by Erno Penzias and Robert W. Wilson. Who predicted its existence is controverted. The "temperature of space" was first estimated to be 5 or 6K by Charles Édouard Guillaume[41] in 1896. In 1926 Sir Arthur Stanley Eddington found an astonishingly precise value[42], 3.18K. But these results were based on star-light considerations and had nothing to do with the expansion of the universe. Predictions based on the latter were made by George Gamow's collaborators Ralf Alpher and George Herman[43], who obtained 5K. None less than Penzias himself avoided discussing who else might be involved in the underlying theories (e.g. the very influential and feared Robert Dicke) by saying: *It is beyond the scope of this contribution to weigh the various theoretical explanations of the 3K* [the MWBR][44]. I follow suit.

Going even further back in time, at an age of the Universe of a few minutes and a temperature of about one billion degrees the preexisting protons (p) and neutrons (n) coalesced into the nuclei of the primordial elements, an idea proposed by Gamow in 1946[45], and subsequently developed in his article with Alpher[46]. Hans Bethe was added to the authors to complete the $\alpha\beta\gamma$ collaboration. This rather weak joke was quite successful and, as you see, it is too often repeated.

The primordial elements are $^4$He, "Helium four" whose nucleus is (2p;2n), Deuterium (p;n); $^3$He (2p;n), "Helium three; and $^7$Li (3p; 4n)[47]. Others, such as Tritium (p;2n) were also made but are not stable. Free neutrons surviving primordial nucleosynthesis also decay. The average abundances of primordial elements in the Universe have changed since they were made. Stars produce $^4$He and destroy $^7$Li. Consequently, the extraction of the primordial element fractions is not so simple. Yet, the data agree with the predictions[45], as shown in Figure 5. The agreement is perfect (but for $^7$Li), considering in particular that the ratios extend over nine orders of magnitude.

---

[39] Cosmologists call this epoch "recombination". Electrons and atomic nuclei had never been combined before, but cosmologists favor confusing language, competing in this with particle physicists.

[40] In cosmology the subindex 0 denotes "now": the beginning of the future, not the beginning of the past.

[41] C.-E. Guillaume, La Nature **24**, series 2, p. 234 (1896).

[42] A. S. Eddington, *The Internal Constitution of the Stars*, Cambridge University Press, Chapter 13, p. 371 (1988). Reprint of the 1926 edition.

[43] R. A. Alpher, and R. Herman, Nature **162**, 774 (1948); Phys. Rev. **75**, 1089 (1949).

[44] A. A. Penzias, in: Cosmology, Fusion & Other Matters, F. Reines (ed.), Colorado Associated University Press, 29-47 (1972).

[45] G. Gamow, Phys. Rev. **70**, 527 (1946). For recent results see https://pdg.lbl.gov/2020/reviews/rpp2020-rev-bbang-nucleosynthesis.pdf

[46] R. A. Alpher, H. Bethe & G. Gamow, Phys. Rev. **73**, 803 (1948).

[47] The processes combining protons and neutrons into light nuclei are simple and well understood, but depend on many variables and require computer "codes" to obtain the results. The first, for $^4$He, was written by Phillip James (Jim) Edwin Peebles. An early complete code is that of R. V. Wagoner, R. C. Herman and F. Hoyle, Ap. J. **148**, 3 (1967).

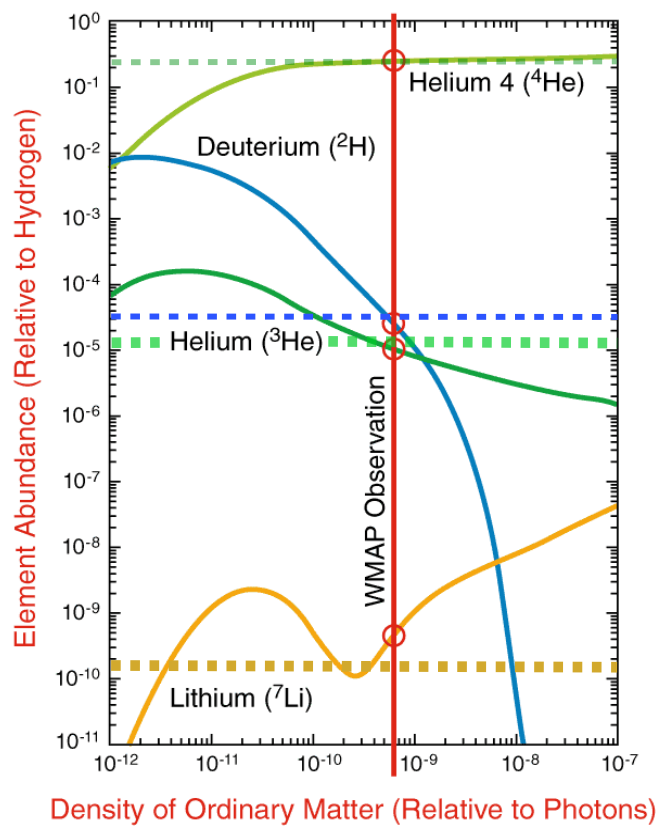

Figure 5. Abundances of the primordial elements as functions of the ratio, $\eta$, of the number of protons plus neutrons to MWBR photons. The circles are the nucleosynthesis predictions[45] for the measured value of $\eta$. The dotted lines are the observed abundances. © NASA.

The abscissa of Figure 5 is, literally, the crux of the matter. It spans values of $\eta \equiv n_B/n_\gamma$, the universal average baryon number density (protons plus neutrons per unit volume) divided by the number density of photons, previously defined. The continuous lines in the figure are the predictions. The dashed horizontal lines are the measured abundances of the different elements, relative to hydrogen. Lo and behold, there is a value of the abscissa, $\eta \cong 6 \times 10^{-10}$, for which theory and observations precisely agree, but for $^7$Li, the not too worrisome fly in the ointment[48]. Even more satisfactorily, this value of $\eta$ agrees with independent measurements of the amount of ordinary matter in the universe, by the WMAP satellite (as in the figure), and by other satellites (Planck) or ground-based microwave antennas[49].

---

[48] One should have expected to have $\eta = 0$, since a universe with as much matter as antimatter is easier to explain, though it would not contain anybody to explain anything. The visible universe might have been made of very large separate matter and antimatter domains. But we know it is not, see Andrew J. Cohen, Álvaro De Rújula and Sheldon L. Glashow, Astroph. J. **495**, 539 (1998).

[49] The consistency of the results is also improving thanks to recent measurements of the reaction proton + deuteron (pn) → triton (ppn) + photon, by the LUNA collaboration, V. Mossa et al., Nature **587**, 210 (2020).

Half hidden in the previous paragraph there is a surprising fact: the measured amount of ordinary matter is only about 1/6 of what would be needed to account for all of matter, including dark matter. In other words, the ordinary matter present in the universe when it was minutes old is insufficient for it to have evolved into something rather non-luminous, such as "Jupiters", failed stars or what not. And this brings us to look at the entire universe more explicitly.

# The ΛCDM model

Particle physicists have a *Standard Model*. It is not a model, but a theory with an extremely long list of correct predictions. Easy, you might say --if you happen to be a cosmologist-- since the model has some 17 parameters that must be taken from the observations. This is an unfair criticism, for the model entails very many predictions of impressive precision that depend on a small subset of the parameters (a two-parameter example is the magnetic moment of the electron, a couple of extra parameters and you get all of atomic physics and chemistry).

Cosmologists also have a benchmark: the *ΛCDM model*, with 6 or 7 parameters to be fit to observations. Incorporating a lot of information from the particle-physics model (of all forces but gravity), the ΛCDM model is also strikingly successful.

The Λ in ΛCDM is the currently most fashionable entity: Einstein's **cosmological constant**. Serious authors dislike information originating from a single source. The sempiternally repeated story that Einstein considered Λ to be his "biggest blunder" apparently has just one source: George Gamow[50]. It is thus to be doubted[51] and not even mentioned.

The cosmological constant is interpreted since the 1960's as the energy density of the vacuum, a notion first proposed (in writing) by Erast Borisovich Gliner[52] and intensely analysed by Zel'dovitch. The energy density of *the vacuum!* **The vacuum is not empty!!** Physicists accept such a weird and fearless notion because they distinguish between the vacuum (a physical reality) and nothingness (an abstract notion). Between *Le vide et le néant*, to put it more elegantly.

What does the rest of the ΛCDM acronym stand for? **Cold Dark Matter**. "Cold" means that the "particles" of dark matter, elementary or not, are non-relativistic in the current stage of the evolution of the universe. "Non-relativistic" means that their velocities are small, compared with the velocity of light. The CDM notion must once have been urgently looking for authors, for it was introduced thrice in 1982: by J. Richard Bond, Alex Szalay and Michael Turner[53], James Peebles[54], and George Blumenthal, Heinz Pagels and Joel Primack[55]. Comparing the titles of their papers is quite informative. In its most strict definition, CDM interacts only gravitationally, even with itself.

---

[50] G. Gamow, *My Worldline*, (Viking Press) 1970, p44.

[51] Livio, M. *Brilliant Blunders* (Simon&Schuster) p233 (2013).

[52] Erast Borisovich Gliner, ZhETF49 542 (1965). Wolfgang Pauli may have been the first to consider the gravitational effects of vacuum fluctuations, see Helge Kragh, History of Exact Sciences 66, 199 (2012).

[53] J. R. Bond, A. S. Szalay, M. S. Turner, *Formation of galaxies in a gravitino-dominated universe*. Physical Review Letters. **48** (23): 1636–1639 (1982).

[54] P. J. E. Peebles, *Large-scale background temperature and mass fluctuations due to scale-invariant primeval perturbations*. The Astrophysical Journal. **263**: L1 (1982).

[55] George R. Blumenthal, Heinz Pagels and Joel R. Primack, *Galaxy formation by dissipationless particles heavier than neutrinos*. Nature. **299** (5878): 37–38 (1982).

The $\Lambda$CDM model is based on the Friedman-Lemaître-Roberson-Walker solutions of Einstein's equations of General Relativity, his theory of space, time, gravity and its source[56]. The solutions describing the universe are obtained assuming a **cosmological principle**, the contention that over large enough domains the universe is homogeneous and isotropic. Nathan Secrest and collaborators argue that this principle may be wrong on relevant distance scales[57]. If so, $\Lambda$ might indeed be Einstein's greatest misconception.

The main success of the $\Lambda$CDM model concerns the analysis of the *anisotropies* of the CBR. The all-sky map[58] of the temperature of this radiation, reaching us after having travelled since recombination for some 13 billion years, is perfectly dull: a homogeneous image at a temperature of about 2.7K in all directions. But once one subtracts the mean temperature signal to look for anisotropies at the $10^{-3}$ level, the result is that of the upper panel of Figure 6. It shows a dipolar pattern and an "equatorial" strip. The "dipole" signal is interpreted as being due to our motion, at a velocity $\sim 10^{-3}$ of that of light, relative to a "universal" rest system in which the CBR would be maximally isotropic[59]. The equatorial strip is our galaxy shining in interstellar light, as first analysed long ago by Guillaume and Eddington, whom we mentioned.

The lower panel shows temperature fluctuations of size $\sim 10^{-5}\, T_0$, obtained after subtraction of the dipole and of the galaxy in which we are immersed, which is a "foreground" to be eliminated like crashed insects on a car's windscreen.

---

[56] The source of the gravitational field in general relativity is the "stress-energy-momentum tensor" of anything (including a gravitational field) for which at least one of the cited entities does not vanish.

[57] For a recent report, see *A Test of the Cosmological Principle with Quasars*; Nathan Secrest, Sebastian von Hausegger, Mohamed Rameez, Roya Mohayaee, Subir Sarkar, Jacques Colin, https://arxiv.org/abs/2009.14826

[58] A Mollweide projection: an equal-area, elliptical projection used for maps of the world or the night sky.

[59] Running in warm still air, one also feels hotter (more abundant and energetic) air molecules hitting one's face.

papers is quite informative. In its most strict definition, CDM interacts only gravitationally, even with itself.

The $\Lambda$CDM model is based on the Friedman-Lemaître-Roberson-Walker solutions of Einstein's equations of General Relativity, his theory of space, time, gravity and its source[56]. The solutions describing the universe are obtained assuming a **cosmological principle**, the contention that over large enough domains the universe is homogeneous and isotropic. Nathan Secrest and collaborators argue that this principle may be wrong on relevant distance scales[57]. If so, $\Lambda$ might indeed be Einstein's greatest misconception.

The main success of the $\Lambda$CDM model concerns the analysis of the *anisotropies* of the CBR. The all-sky map[58] of the temperature of this radiation, reaching us after having travelled since recombination for some 13 billion years, is perfectly dull: a homogeneous image at a temperature of about 2.7K in all directions. But once one subtracts the mean temperature signal to look for anisotropies at the $10^{-3}$ level, the result is that of the upper panel of Figure 6. It shows a dipolar pattern and an "equatorial" strip. The "dipole" signal is interpreted as being due to our motion, at a velocity $\sim 10^{-3}$ of that of light, relative to a "universal" rest system in which the CBR would be maximally isotropic[59]. The equatorial strip is our galaxy shining in interstellar light, as first analysed long ago by Guillaume and Eddington, whom we mentioned.

The lower panel shows temperature fluctuations of size $\sim 10^{-5}\, T_0$, obtained after subtraction of the dipole and of the galaxy in which we are immersed, which is a "foreground" to be eliminated like crashed insects on a car's windscreen.

---

[56] The source of the gravitational field in general relativity is the "stress-energy-momentum tensor" of anything (including a gravitational field) for which at least one of the cited entities does not vanish.

[57] For a recent report, see *A Test of the Cosmological Principle with Quasars*; Nathan Secrest, Sebastian von Hausegger, Mohamed Rameez, Roya Mohayaee, Subir Sarkar, Jacques Colin, https://arxiv.org/abs/2009.14826

[58] A Mollweide projection: an equal-area, elliptical projection used for maps of the world or the night sky.

[59] Running in warm still air, one also feels hotter (more abundant and energetic) air molecules hitting one's face.

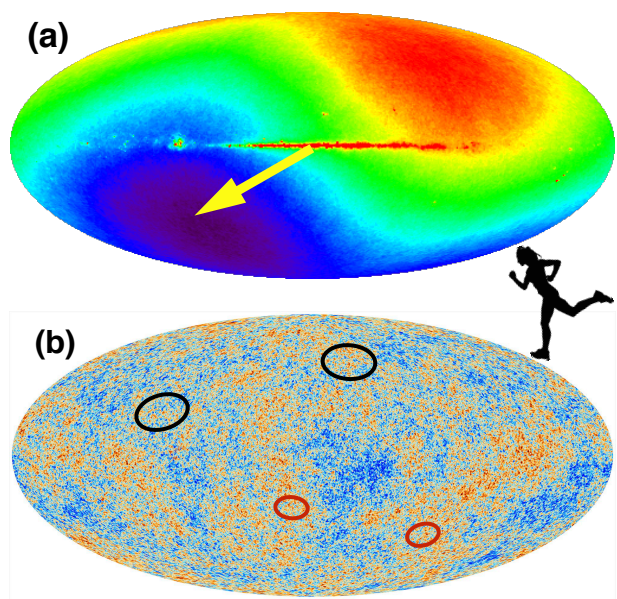

Figure 6. Maps of the irregularities of the CBR, color-coded for human eyes. In (a) the dipole is dominant and the galaxy is visible © NASA/WMAP. In (b) they are subtracted, revealing the true intrinsic CBR anisotropies, © ESA and Planck Collaboration. The anisotropies are analyzed in regions of various angular apertures, or "scales", of which two examples, each in two directions, are shown. Runner: https://openclipart.org/detail/259731/female-runner-2

Your ears and your brain are truly incredible. If you listen to a note played by a musical instrument, you may immediately conclude that it is, e.g., a flute. You may even specify whether it is a wooden or a metal one. It all hinges on the relative magnitudes of the main note and its harmonics. The analysis of the results in the lower panel of Figure 6 is similar. By plotting the square of the temperature fluctuations as a function of the angular scale of the spots one is looking at, summed over all directions in the sky, one finds a series of *acoustic peaks*: a main harmonic at about 1° and six harmonics at smaller angles (~1/2°, ~1/3°, …), as shown in Figure 7.

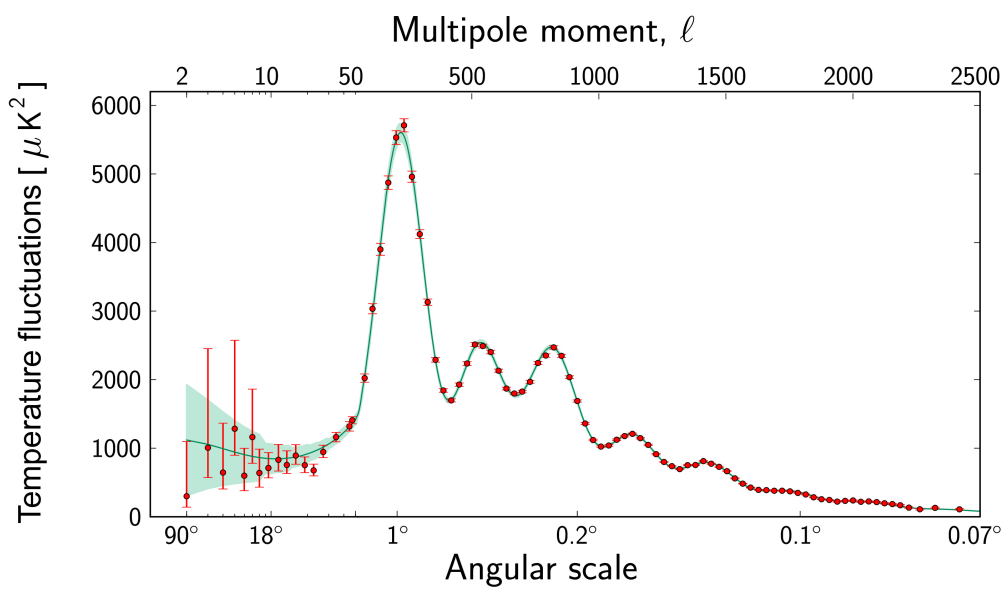

Figure 7. Angular (or multipolar) analysis of the CBR. © ESA/Planck satellite.

The acoustic peaks in the angular analysis of the CBR are due to pre-existing inhomogeneities of the density of ordinary and dark matter. They were predicted, well before inflation or the ΛCDM model became the standard lore, by the usual "Russian suspects": Andrei Sakharov, Rashid Sunyaev and Zel'dovich[60] and, in the West, by Edward Harrison[61], Jim Peebles and J. T. Yu[62]. A discussion of the allegedly inflationary origin of these non-uniformities would take us far too far. Suffice it to say that they are supposed to be due to quantum fluctuations, so that not only the smallest objects (such as subatomic particles) but also the largest ones (the inhomogeneities of the matter and radiation densities at recombination) require quantum mechanics to be understood.

Lo!, from the sizes and positions of acoustic peaks one can decide not only which musical instrument emitted a note, but also what it is that rings in the early universe. That is, what it is made of. The surprising result is shown in Figure 8. The *dark energy* (which we have been calling the vacuum energy or the cosmological constant) is the largest contributor to the energy density of the universe (~70% of the total). Dark matter is next (~25%). Ordinary matter is the miserable remaining ~5%. That these fractions are currently not so different is an unexplained coincidence.

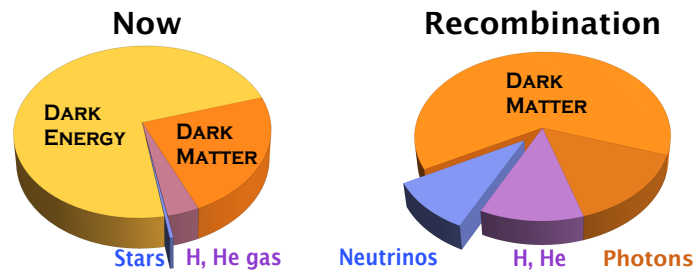

Figure 8. Pie-plots of the contributions of the various "ingredients" of the universe to its average energy density, normalized to the total one. Left $t = t_0$. Right $t = t_r$.

The fractions in Figure 8 change with time as the universe expands. The right hand side shows them at $t = t_r$. The differences relative to $t = t_0$ are due to how energy densities vary as the universe expands. For non-relativistic ordinary and dark matter $E = m c^2$ and, $m$ being constant, the energy density varies with the inverse of the expanding volume. For photons $E = p c$ and their momentum, $p = h \nu/c$, gets red-shifted by an extra inverse size. Nearly massless neutrinos behave like photons. Only a cosmological constant, high time we get to the meaning of its name, stays… constant (and starts to be dominant only at $t \gg t_r$).

In our understanding of the universe based on General Relativity there is a critical density, $\rho_c$, for which the geometry of three-dimensional space is flat (Euclidean), not curved, as the two-dimensional surface of the Earth is. The ratio, $\Omega \equiv \rho_{TOT}/\rho_c$, of the observed total density (summed over all ingredients) to the critical one is observed to be $\Omega = 1$, with better than 1% precision[63]. That is not quite a

---

surprise[64] and it is a blessing of sorts. A minimally over-critical universe at, say, the time of primordial nucleosynthesis (an epoch that we understand) would have ended, eons ago, in a big crunch. A slightly under-critical one would have become so diluted that galaxies, stars and planets would have never been born. In the inflationary paradigm $\Omega = 1$, with enormous precision, is a prediction.

The $\Lambda$CDM model is much more than a 6-parameter tool to fit the data[65]. Some of its successful implications are predictions[61] that preceded it. For example the *baryon acoustic oscillations,* BAOs. Primordial non-homogeneities are supposed to be *adiabatic:* the same for dark matter, baryons, electrons and photons. Dark matter is only subject to gravitational forces, the other species constitute a plasma that, unlike dark matter, is reactive to pressure. In this plasma an over-density generates an outgoing "sound" wave that leaves an imprint, also an over-density, in the distribution of ordinary matter after recombination. The statistical study of the two-galaxy correlation function reveals such a feature[66], precisely as expected.

The different parts of any decent endeavour, it goes without saying, should not disagree with each other. Yet, some authors grandly refer to the results in Figure 9 as *The Cosmic Concordance.* Notice how the combination of different measurements zooms in to select a flat universe dominated by its vacuum and total matter energy-densities.

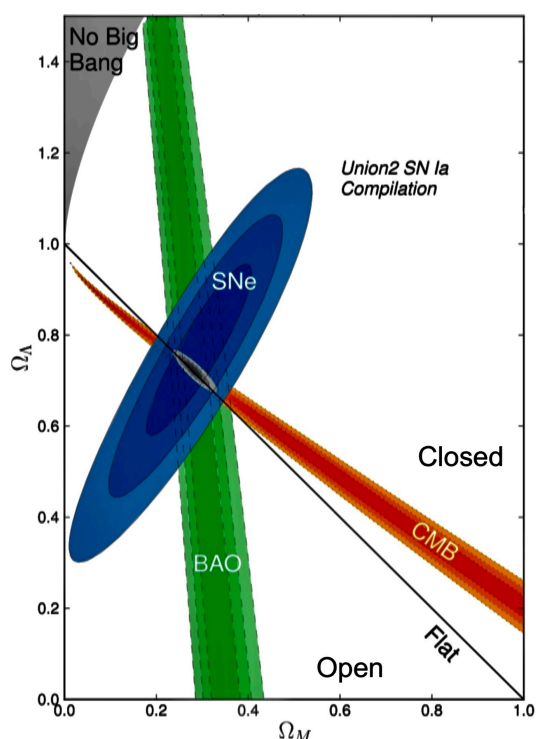

Figure 9. The combined results from the analyses of the Cosmic Microwave Background, distant Supernovae of Type Ia and Baryon Acoustic Oscillations in the plane $\Omega_M \equiv \rho_M/\rho_c$ vs $\Omega_\Lambda \equiv \rho_\Lambda/\rho_c$ of matter and vacuum energy densities, normalized to the critical density.

---

https://pdg.lbl.gov/2020/reviews/rpp2020-rev-astrophysical-constants.pdf

[64] Before a non-zero $\Lambda$ was established, $\Omega = 1$ was the (artificial) boundary between ever expanding and eventually crushing universes, as a not-so-young reader may once have learned.

[65] As John Von Neumann supposedly said: *With four parameters I can fit an elephant, and with five I can make him wiggle his trunk.* The French version, attributed to Jacques Hadamard, requires 100 parameters.

[66] M. Colless et al. (2dFGRS team), Mon. Not. R. Astron. Soc. **328**, 1039 (2001) and http://arxiv.org/abs/astro-ph/0306581; D. G. York et al. (SDSS collaboration), Astron. J. **120**, 1579 (2000).

The *polarization* of the CBR, discovered by DASI[67], is another success story. There are two types of CBR polarization: *E-modes* and *B-modes*[68], induced by the last photon-electron scatterings prior to the detection, here, of the CBR photons. The *E* and *B* polarization anisotropies are correlated with the temperature ones (*T*). So far, only the results on *E*-modes are beyond doubt. On occasion Wikipedia is perfect. To quote it verbatim regarding the results of the Planck collaboration[69]: [there are] *six peaks in the temperature-polarization (TE) cross spectrum, and five peaks in the polarization (EE) spectrum. The six free parameters can be well constrained by the TT spectrum alone, and then the TE and EE spectra can be predicted theoretically to few-percent precision with no further adjustments allowed: comparison of theory and observations shows an excellent match.*

The ΛCDM model is reputedly robust[70]. The values of some of the parameters it implies --for instance the Hubble constant-- are hardly modified by adding parameters to it, such as a non-zero space curvature. Treated as a free parameter, the "effective" number of (light) neutrinos is measured[63] to be $N_\nu = 2.99 \pm 0.17$, in agreement with --but weaker than-- the recently improved result[71] from LEP, $N_\nu = 2.9963 \pm 0.0074$. As for the sum of the three known neutrino masses, the combined BAO plus ΛCDM result[63], $\Sigma\, m_\nu < 0.12$ eV (95% c.l.), is better than the best limits on the electron neutrino all by itself and orders of magnitude more strict than the limits including the muon and tau neutrinos. Quite a feat.

Suppose that, prior to recombination, when atoms could not yet form, you were to choose your favourite proton and your favourite electron. The distance between them would be increasing with time as the universe expands. But after recombination a hydrogen atom has a fixed size, it is a "ruler" with which you can determine that the distance between two distant atoms increases with time[72]. At recombination atoms *separate* from the expansion. More complex structures: overdensities of dark matter, clouds of ordinary matter, stars, galaxies and clusters separate at later times. With the help of the ΛCDM model, this "generation of structures" can be studied in detail, currently with increasing effort and success[73].

**AND, AFTER ALL, WHAT IS Λ?**

First, a bit technically. Einstein defined his cosmological constant as a term in his equations describing the field of gravity, $g_{\mu\nu}$, which is also the four by four "metric" matrix describing space-time[74], the cosmological constant is an additional source of gravity of the form $\Lambda\, g_{\mu\nu}$. Its effect on the evolution with time of a universe obeying the cosmological principle is reflected in the "first Friedman equation":

$$3\left(\frac{da/dt + k\,c^2}{a}\right)^2 = 8\,\pi\,G_N\,\rho + \Lambda\,c^2$$

---

[67] J. M. Kovac, E. M. Leitch, C. Pryke, J. E.; Carlstrom, N. W. Halverson, W. L. Holzapfel, *Detection of polarization in the cosmic microwave background using DASI*. Nature. **420** (6917): 772–787 (2002). arXiv:astro-ph/0209478.

[68] U. Seljak and M. Zaldarriaga, Phys. Rev. Lett., **78**, 2054 (1997); M. Kamionkowski, A. Kosowsky and A. Stebbins, Phys. Rev. Lett., **78**, 2058 (1997).

[69] Planck Collaboration (2016). *Planck 2015 Results. XIII. Cosmological Parameters.* Astronomy & Astrophysics. **594** (13): A13, arXiv:1502.01589

[70] "Robust" is an ugly adjective often abused by some scientists to describe results that are not.

[71] Patrick Janot and Stanisław Jadach, Phys. Lett. **B803** 135319 (2020).

[72] Prior to recombination the size of protons may be the ruler. At earlier times when only elementary particles were around, anything that varies, such as the rate of some reaction (and the speed of light) could be your clock and ruler. All one needs to specify times and distances is something to be "happening".

[73] For a relatively recent review, see M. Vogelsberger, F. Marinacci, P. Torrey, et al. *Cosmological simulations of galaxy formation.* Nat. Rev. Phys. **2,** 42 (2020).

In it, $a(t)$ is a distance scale (between, for instance, two very far apart galaxies), $G_N$ is Newton's constant and $\rho$ is the energy density in matter and radiation. In the convention that $a = 1$ now, $k = 1, 0, -1$ describes the 3D space curvature, respectively closed, flat or open.

If the $\Lambda$ term dominates over the $\rho$ and $k$ terms the solution is of the form $a \propto \exp\left[c\,t\,\sqrt{\Lambda/3}\right]$ and, exponentially with time, space "inflates", making its 3D curvature tends to zero. In the far future distant galaxies will be unobservable and cosmology a chapter of ancient history.

The cosmological constant can be regarded as a gravitational "force" between separate points in space. Newton's gravitational force "derives from a potential" inversely proportional to distance, $V_N \propto 1/r$. The analogous "potential" for the $\Lambda$ term is linear in distance, $V_\Lambda \propto r$. That is why, though $\Lambda$ dominates the current universe, its effects on "small" scales, such as a laboratory or even the solar system, are far too tiny to be observable.

More precisely, the cosmological constant is a maximum negative pressure, $w = -1$ in the *equation of state* relating the pressure, $p$, and density, $\rho$, of the vacuum, as sources of a gravitational field: $p = w\rho c^2$.

In the theory of an inflationary universe, for a brief while after $t = 0$, the vacuum energy density of an "inflaton" spin-zero field plays the role of a cosmological constant. But it is not constant in time, its energy density evolves into that of the other substances[75]. The currently measured cosmological constant could also be time-varying. Such a possibility may solve the Hubble constant "tension", if indeed there is one.

What $\Lambda$ is, most seriously, is a Godzilla-sized fly in the ointment. Now that it has been measured, one can no longer entertain the hope that some still undiscovered dynamical or symmetry principle will set it to zero. The measured value of $\Lambda$ approximately corresponds to the mass-energy of three and a half hydrogen atoms per cubic meter, that is $\sim 5 \times 10^{-10}$ Joules/m$^3$. Somehow extracting its energy from a cubic meter of this stuff you could light a 100 Watt lamp for a few billionths of a second. Yet, this locally tiny value of $\Lambda$ dominates the universe at its large scales.

In a hypothetical self-respecting relativistic quantum theory of gravity, $\Lambda$, which is an energy per unit volume, should only depend on the gravitational parameter of the theory, $G_N$, the speed of light, $c$, and the reduced Planck constant, $\hbar$. Explicitly, on dimensional grounds, one would expect Λ to be of the order of $c^3/(\hbar\, G_N)$. This is more than 122 orders of magnitude larger than the observed Λ. The worst estimate ever, with no known generally accepted explanation.

---

[74] For a non-curved space-time one can choose $g_{00} = 1, g_{11} = g_{22} = g_{33} = -1$, with the other 12 entries vanishing.

[75] A process called "reheating", though, supposedly, nothing had been hot before.

Steven Weinberg, in 1987, found an upper bound to the value of Λ, beyond which matter would be diluted so fast as to prevent galaxy formation and life[76]. This would diminish the size of the conundrum by some 120 orders of magnitude, still not enough. (Not quite) similar *anthropic principles* are very much in vogue, along with *multiverses*[77]. We the old-fashioned shall take these ideas more seriously as soon as they help make a correct prediction.

**DISCORDANCE?**

By itself, the ΛCDM model would be unable to explain the observations. The main conundrum is the near-perfect homogeneity of the MWBR temperature measured in different directions: regions at scales larger than a few degrees require absurdly tuned initial conditions, for they should never have been in causal contact in their past. An inflationary universe[78] --invented before the ΛCDM model to solve other problems[79]-- takes the "absurd" adjective away, replacing it by a possible understanding of the near-homogeneity of the CBR and, as a bonus, the prediction that the universe is 3D flat.

As new data are collected, the inflationary ΛCDM model begins to face problems, which Leandros Perivolaropoulos and Foteine Skara[17] and Eleonora Di Valentino, Alessandro Melchiorri, and Joseph Silk[80] have recently summarized and systematically analized[81]. Some of their results are shown in Figure 10. Its (b) panel is the $(w, H_0)$ plane, the (a) one is the $(H_0, \Omega_k)$ plane, with $\Omega_k$ the space curvature, normalized to the critical density, and playing a role of an extra $\rho_k \propto a^{-2}$ density in Friedman's equation.

The data in Figure 9 and the ones labelled "Planck + BAO" in Figure 10 very snuggly agree with an *assumed* 3D-flat, $\Omega_k = 0$ universe. The addition of $\Omega_k$ as a parameter to be fit, it is argued, results in the rest of the contours in Figure 10a, significantly advocating for a closed universe[81]. If the $w = 1$ assumption of the ΛCDM model is abandoned in favour of an extra $w$ parameter, a cosmological constant is similarly disfavoured[81], as in Figure 10b.

---

[76] Steven Weinberg, Phys. Rev. Lett. **59**, 2607 (1987).

[77] An early defender of such views was Giordano Bruno. To quote him again: *Stimo infatti cosa indegna della infinita potenza divina che, potendo creare oltre a questo mondo un altro e altri ancora, infiniti, ne avesse prodotto uno solo, finito.* That is: I find unbecoming to the infinite divine might that, being able to create beyond this world another one and others yet, infinitely many, it [the divine might] would have created only one, a finite one.

[78] The original authors are Alexei Starobinsky, Allan Guth, Andrei Linde, Paul Steinhardt, Viatcheslav Mukhanov, and others. The paternity of this idea is shared and hard to weigh.

[79] For a more detailed elementary discussion of cosmology and the inflationary paradigm, see, e.g. Alvaro De Rújula, *Enjoy your Universe / you have no other choice.* Oxford University Press, 2018.

[80] E. Di Valentino, A. Melchiorri, and J. Silk, Phys. Lett. B, 761, 242, (2016), Nature Astron., **4**, 196, (2019).
   N. Aghanim, et al. 2020a, Astron. Astrophys., **641**, A6, (2020); Astron. Astrophys., **641**, A5, (2020).

[81] E. Di Valentino, A. Melchiorri and J. Silk, arXiv:2003.04935v2.

22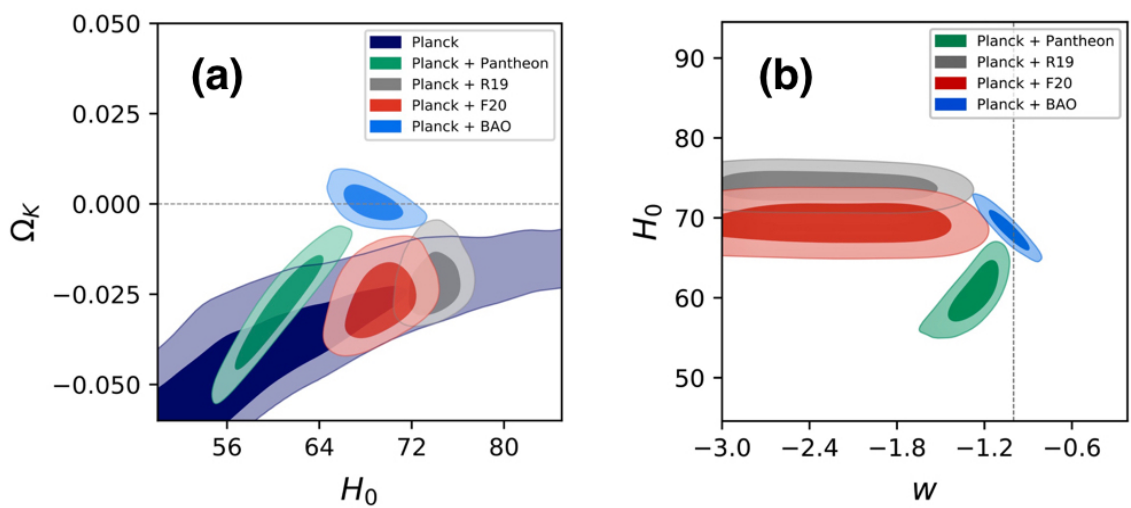

Figure 10. The Hubble constant versus the curvature density $\Omega_k$ (a), and the equation of state parameter $w$ (b)[81]. The recent data employed in this figure are labeled Pantheon[82], R19[83], and F20[84]. The contours are 68% and 95% confidence levels[81]. The dotted lines are the ΛCDM hypothesis.

Particle physicists adamantly look for new physics *beyond* their Standard Model, whose extreme success in confronting the data is not often controverted. But cosmology is not similarly uncontroversial. If the data analysis resulting in Figure 9 stands the critique from the ΛCDM advocates, we would be entering an era of *Cosmic Discordance.* And, if so, cosmologists would have to look for new physics beyond, or perhaps *within,* their standard model.

**WHAT IS THE DARK MATTER?**

Consulting the not peer-reviewed arXiv.org website --where physicists, mathematicians, epidemiologists and so many other scientists post their "preprints"-- one verifies that there are, every workday, several new theories of what dark matter may be and, almost that frequently, proposals of new ways to detect it. While the proposed experiments are very often extraordinarily ingenious, the task of the theorists may be impossible: imagine dark matter to be as complex as ordinary matter is, but not too similar; who would be capable of inventing a whole new world? The only hope for a right guess on the nature of dark matter is that what we know about ordinary matter gives us a good hint. Black holes are an example. I shall only discuss two others.

*Axions*

It is frequently stated that, after the discovery of the Higgs boson, the standard model of particle physics is "complete". Fake news. Quantum Chromo-Dynamics (QCD) --the section of the theory dealing with the "strong" interactions of quarks and gluons-- has a problem. The combined symmetry of charge conjugation and parity (CP) is slightly "violated" by the weak interactions, something that is very well understood. But QCD also turns out to violate CP, at a strong level excluded

---

[82] D. Scolnic et al., Astrophys. J., **859** (2018).
[83] A.G. Riess, Nature Rev. Phys., **2**, 10 (2019).
[84] W.L. Freedman, B. F. Madore, T. Hoyt, et al. 2020, doi: 10.3847/1538-4357/ab7339.

by experiment.

To cure the "strong CP problem" Roberto Peccei and Helen Quinn imposed an extra symmetry on the theory that, combined with the highly non-trivial nature of the QCD vacuum, solved the problem[85]. Physicists and their eternal fencing with the vacuum, you'll correctly say! Frank Wilczek and Steven Weinberg independently realized that the extra symmetry implied the existence of a new spin-zero particle, that Wilczek called the *axion*[86].

The "natural" energy scale where to place the dynamics of the axion was the scale of the weak interactions, some 250 GeV. The estimated mass of the axion, inversely proportional to that scale, was in the keV domain, and the axion couplings to other particles, though proportional to the small axion mass, were strong enough for the particle to be swiftly discovered. It was not. Arguments based on "naturalness", whatever that is, tend to end that way.

Totally undeterred, theorists increased the axion scale, inventing "invisible" axions. Insensitive to the theorists' language, experimentalists set to look for them. Meanwhile and just in case, theorists came up with Axion-Like-Particles (ALPs), for which the mass and the strength of the couplings to photons ($\gamma$) and to ordinary matter are unrelated. A very characteristic coupling of an axion field, $a$, is the one to two photons: $\mathcal{L}_{a\gamma\gamma} = g_{a\gamma\gamma}\, a\, \vec{E} \cdot \vec{B}$, describing, for instance, the slow $a \to \gamma\,\gamma$ axion decay ($\vec{E}$ and $\vec{B}$ are the electric and magnetic parts of the photons' fields, which may be "real" as in light, or "virtual", as for a static $\vec{E}$ or $\vec{B}$ field).

The most secure way to look for ALPS or axions is to make and detect them "by yourself". One way, called LSW (Light Shining through Walls), is to have laser light, $\gamma$, travel in a magnetic field, $\vec{B}$, to generate axions ($\gamma + \vec{B} \to a$), "shine" the weakly interacting axions through a wall and attempt to detect them on the other side of it via the inverse process ($a + \vec{B} \to \gamma$), a technique twice invoking the coupling $\mathcal{L}_{a\gamma\gamma}$. Though we are mainly interested in axions as dark matter, I could not resist referring to this particular approach, if only because of the grandeur of the endeavor[87]. The domain excluded by a first generation ALPS experiment, the reach of ALPS II and various other results are shown in Figure 11.

---

[85] It is highly non-trivial to explain this subject in detail. One cannot do it briefly, as I recently learned, having failed in A. De Rújula, *Aperitivos* [tapas] *de materia oscura*. Revista Española de Física, 34, 1 (2020).

[86] For a brief (50 page) review of the strong CP problem and the relevant literature, see Roberto Peccei, arXiv:hep-ph/9807516.

[87] The experiment Any Light Particle Search II (ALPS II) in the ex-high-energy laboratory Deutsches Elektronen-Synchotron (Hamburg) measures hundreds of meters and uses dozens of superconducting magnets. See https://arxiv.org/pdf/1611.05863.pdf.

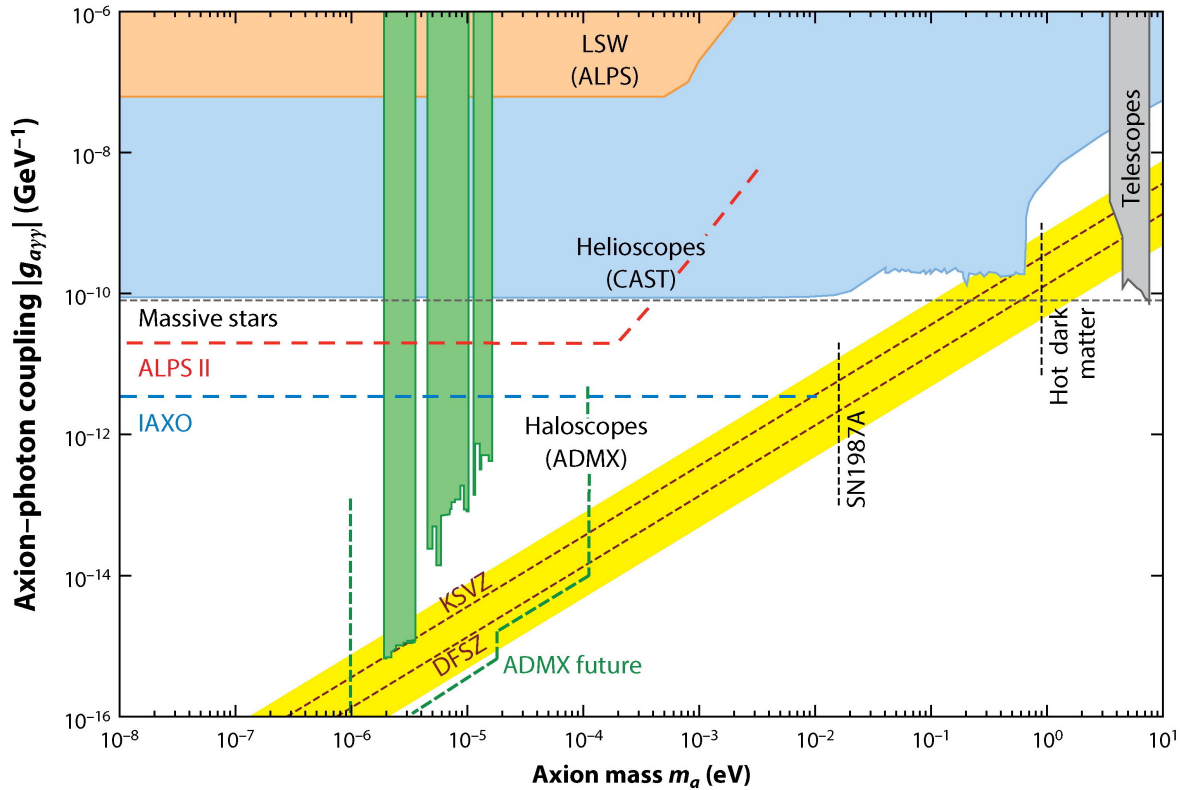

Figure 11. The plane of axion mass vs its two-photon coupling. The yellow band is the expectation in two types of invisible axion models. The colored dashed lines reflect the reach of future experiments. For more details, see the text and the comments to the original figure in its source: https://pdg.lbl.gov/2019/reviews/rpp2019-rev-axions.pdf

*Helioscopes* are devices, invented by Pierre Sikivie, to look for axions made in the sun by photons interacting in its interior with the electric field of an atomic nucleus ($\gamma + \vec{E} \to a$). In an electromagnetic resonant cavity permeated by a magnetic field the solar axions would be reconverted into an observable X-ray photon ($a + \vec{B} \to \gamma$, once more). The experiment CAST (CERN Axion Solar Telescope) used a magnet originally intended for the LHC (Large Hadron Collider) as an "axioscope", obtaining the limits of Figure 11. IAXO is its future incarnation.

There is an extensive literature on the various ways in which axions could be made in the early universe[88] and on the possibility that they are a form of cold dark matter or, contrariwise, a hot relic[89], like the hard-to-detect fossil neutrinos are convincingly argued to be. Cosmic axions decaying into photons ($a \to \gamma\gamma$) have not been seen. That excludes the domain labeled "Telescopes" in Figure 11.

*Haloscopes*, another brain child of Pierre Sikivie, are high-Q microwave cavities

---

[88] J. Preskill, M. B. Wise, and F. Wilczek, *Cosmology of the Invisible Axion,* Phys. Lett. B120, 127 (1983).

L. Abbott and P. Sikivie, *A Cosmological Bound on the Invisible Axion,* Phys. Lett. B120, 133 (1983).

The Not So Harmless Axion, Michael Dine and Willy Fischler, Phys.Lett. B120, (1983) 137.

[89] Jeff A. Dror, Hitoshi Murayama, Nicholas L. Rodd, *The Cosmic Axion Background,* arXiv:2101.09287.

permeated by a magnetic field. They would convert all the energy of axions of mass $m_a$, bound with velocities $v/c \sim 10^{-3}$ in the galactic halo, into a signal of frequency $v = m_a/h$, of width $\sim(v/c)^2$. Assuming that axions constitute the bulk of the dark matter of the galaxy and very very patiently tuning the frequency, the green domains in Figure 11 have been excluded. Also shown is the reach of the currently running ADMX haloscope (Axion Dark Matter eXperiment).

*Supersymmetry*

No other possible dark matter candidate is as theoretically "required" as the axion. Yet, it is by no means the most wanted culprit. That honour goes to the Lightest Supersymmetric Particle (LSP) and its generalizations: the Weakly Interacting Massive Particles (WIMPs, to prolong the anthropic name-giving tradition started by Brown Dwarfs and MACHOs).

Supersymmetry[90], also called SUSY, is an extremely beautiful idea. If it were an unbroken local symmetry of nature there would be, for each elementary particle, a *superpartner* with the same mass and half a unit larger or smaller spin. That is not correct, a massless spin-1/2 *photino*, for instance, does not exist. There is no known simple and elegant way to break SUSY, implying that the masses of the many superpartners must be chosen carefully (to avoid immediate trouble) and otherwise arbitrarily, as in the Minimal Supersymmetric Standard Model (MSSM)[91].

If and when the temperature of the primordial plasma was above the masses of the supersymmetric particles, their couplings to themselves and to the standard particles –all of which SUSY specifies— imply that the superpartners were in thermal equilibrium. As the temperature decreased, they must all have decayed into lighter standard particles. Or all but one, the LSP, if there is an unbroken symmetry forbidding the decay of superpartners into only standard particles. Thus, the LSP, if neutral, is a candidate dark matter particle.

The previous syllogism comes with a bonus. If the masses of superpartners are such that they can be discovered at CERN's LHC (recall that their couplings are predicted), then the relic abundances of various LSP/WIMP candidates can be estimated and they turn out to be compatible with the observed dark matter density. This reasoning is very often presented in its inverted form, as a prophecy, based on the existence of the right amounts of dark matter, foretelling the discovery of SUSY at the LHC. The coincidence of these two energy scales is too often dubbed "the WIMP miracle". Not a good omen for scientists to use such language. Indeed, supersymmetry has not been discovered in the advocated "natural" range of energies[92].

It is a never-ending enterprise to prove a theory "right". To prove a theory wrong is quite possible and used to be a source of satisfaction for experimentalists. The LHC experiments have excluded vast regions in the parameter space of tens of theories, supersymmetric or not, some very interesting --like the possible existence of accessible extra space dimensions-- others not so much. This progress is not viewed with the deserved enthusiasm, a sad consequence, perhaps, of having faithfully sold, too early, the super-bear's skin.

---

[90] The history of the original ideas on supersymmetry is not that simple. I have not studied it well enough to attempt to summarize it.

[91] A name implying, in my opinion, an unintended reference to the MSSM's credibility.

[92] Supersymmetry commands very strong feelings, on occasion strong enough for it to be allegedly discovered. For an entertaining recount of such an occurrence, see: *Nobel Dreams: Power, Deceit, and the Ultimate Experiment.* Gary Taubes. Random House, (1987).

**CONCLUSION**

The enormity of the challenge posed on science by the dark side of the universe is unprecedented. Scientists have reacted to it with their characteristic ingenuity and dedication. They have found vast swaths of the land of possibilities to be barren. The search continues. There is no real conclusion.

**Acknowledgements.** I am indebted to Michel Spiro for discussions. I am very indebted to Manuel Aguilar-Benitez, J. Adolfo de Azcárraga, Juan García Bellido, Thibault Damour, Loredana Gastaldo, James Rich and Fabio Truc for their critical reading of the manuscript and/or for many useful comments and suggestions. This project has received funding/support from the European Union's Horizon 2020 research and innovation program under the Marie Sklidowska grant agreement No 860881-HIDDeN.